\documentclass[a4paper]{PoS}
\usepackage[utf8]{inputenc}
\usepackage[T1]{fontenc}
\usepackage{rotating}
\usepackage[sort]{cite}
\usepackage{lineno}

\newcommand{\U}[1]{\,\mathrm{#1}}

\newcommand{\EQG}{E_\mathrm{QG}}

\title{The potential of the HAWC Observatory to observe violations of Lorentz Invariance}

\ShortTitle{HAWC potential to observe Lorentz Invariance Violation}

\author{\speaker{Lukas Nellen}$^a$ for the HAWC collaboration$^b$\\
  \llap{$^a$}I.\ de Ciencias Nucleares, UNAM, México\\
  \llap{$^b$}For a complete author list, see
  \href{http://www.hawc-observatory.org/collaboration/icrc2015.php}{http://www.hawc-observatory.org/collaboration/icrc2015.php}.\\ 
  E-mail: \email{lukas@nucleares.unam.mx}}

\abstract{The framework of relativistic quantum-field theories
  requires Lorentz Invariance. Many theories of quantum gravity, on
  the other hand, include violations of Lorentz Invariance at small
  scales and high energies. This generates a lot of interest in
  establishing limits on such effects, and, if possible, observing them
  directly. Gamma-ray observatories provide a tool to probe parts of
  the parameter space of models of Lorentz Invariance Violation that
  is not accessible in terrestrial laboratories and man-made
  accelerators. Transients, especially gamma-ray bursts, are a
  particularly promising class of events to search for such
  phenomena. By combining cosmological distances with high energy
  emission and short duration, emitting photons up to $30 \U{GeV}$ in less
  than a second, one can measure the energy dependence of the speed of
  photons to one part in $10^{16}$. We will discuss the potential of HAWC
  to detect effects of the violation of Lorentz Invariance and place
  its sensitivity in the context of existing limits.
}

\FullConference{The 34th International Cosmic Ray Conference,\\
		30 July- 6 August, 2015\\
		The Hague, The Netherlands}

\begin{document}

\section{Introduction}
\label{sec:introduction}

The HAWC Observatory \cite{pretz:2015, hawcweb}, located at an
elevation of $4100 \U{m}$ on the volcano Sierra Negra in Mexico,
is expected to be capable of placing limits on certain classes of
models for quantum gravity that predict the violation of Lorentz
Invariance.

Relativistic quantum field theory (RQFT) \cite{peskin-schroeder:1995},
like many modern theories, requires Lorentz Invariance for
consistency. The framework of perturbative RQFT is the basis for the
successful Standard Model of particle physics. On the other hand,
it has not been possible so far to develop a consistent
theory that unifies gravity with RQFTs
\cite{Woodard:2009ns}. Nevertheless, models of Quantum Gravity (QG)
have been developed, many of which predict the violation of Lorentz
Invariance either at very high energies or very short distances
\cite{Smolin:2003rk}. In many cases, this can result in observable
deviations from standard theories \cite{Mattingly:2005re}.

It is typically expected for QG to manifest itself fully at the Planck
Scale, set by the Planck Mass 
$m_\mathrm{pl} = \sqrt{\hbar c / G} \approx 10^{19}\U{GeV}$. 
This makes it also a natural scale at which one expects Lorentz
Invariance to be broken. The Planck Scale is not accessible directly
in the laboratory or in astrophysical objects in the contemporary
universe. It is, 
however, reasonable to expect that small, residual effects of QG could
lead to an observable Lorentz Invariance Violation (LIV) in
astro-particle physics, in particular in observations of high energy
gamma rays. One of the familiar equations that can get modified is the
dispersion relation for photons, resulting in an energy dependent
speed for the propagation of photons, instead of a constant speed
of light in the vacuum $c$. The leading term in the modified
dispersion relation is \cite{Bolmont:2010np}:
\begin{equation}
E^2 \simeq p^2c^2 \Biggl(1 + 
  \xi_n \biggl(\frac{pc}{\EQG}\biggr)^n\Biggr),
\label{eq:dispersion}
\end{equation}
with $\EQG$ being the energy scale for QG effects, normally the Planck scale, 
and $\xi_n$ a dimensionless expansion coefficient. Depending
on details of the model, the leading term is linear ($n=1$) or quadratic
($n=2$). Higher order leading terms are generally not considered.
When quoting limits, it is common to establish them for a combined scale
variable
$\EQG^{(n)} \equiv E_\mathrm{QG} \xi_n^{-1/n}$.
The speed of propagation of photons derived from
eq.~(\ref{eq:dispersion}) is
\begin{equation}
  \label{eq:speed}
  v = \frac{\mathrm{d}E}{\mathrm{d}p} 
  \approx c\Biggl( 1 +
  \frac{n+1}{2} \biggl(\frac{pc}{E_\mathrm{QG}^{(n)}}\biggr)^n \Biggr).
\end{equation}

The speed of photons is no longer constant, which means that photons
emitted simultaneously will arrive at the observer spread over a time
$\Delta t$, which depends on the spread of energy of the photons produced and
the distance to the source. For nearby sources, for example pulsars
in our galaxy, the curvature of the
universe can be neglected when calculating the photon travel time
and one obtains for the time spread
\begin{equation}
  \label{eq:dt-near}
  \Delta t = \frac{(n+1)d}{2c} 
             \frac{\Delta E^n}{\bigl(E_\mathrm{QG}^{(n)}\bigr)^n}
           \approx \frac{(n+1)d}{2c} 
             \frac{E_\mathrm{max}^n}{\bigl(E_\mathrm{QG}^{(n)}\bigr)^n}
\end{equation}
for the distance to source $d$ and $\Delta E^n = E_\mathrm{max}^n -
E_\mathrm{min}^n$. In most cases, the energy range considered spreads
several orders of magnitude and one can approximate $\Delta E^n \approx
E_\mathrm{max}^n$.

For photons traveling cosmological distances, for example those
originating from Active Galactic Nuclei (AGN) or Gamma Ray Bursts
(GRB), one has to take into account the redshift of the source and the
non-trivial metric of the universe. The resulting expression for the
time spread is \cite{Bolmont:2006kk}
\begin{equation}
  \label{eq:dt-far}
  \Delta t = \frac{n+1}{2} H_0^{-1} \frac{E_\mathrm{max}^n}{\bigl(E_\mathrm{QG}^{(n)}\bigr)^n}
\int_0^z \frac{(1+z')^n}{h(z')}\,\mathrm{d}z'
\end{equation}
where $h(z) = \sqrt{\Omega_\Lambda + \Omega_m (1+z)^3}$. The
cosmological parameters used in this paper are
$H_0 = 70 \U{km} \U{s}^{-1} \U{Mpc}^{-1}$,
$\Omega_\Lambda = 0.7$, and $\Omega_m = 0.3$. 
Using the most recent results by Planck \cite{Ade:2015xua} does not
affect the HAWC reference scenarios significantly.

\section{Limits on the Violation of Lorentz Invariance}
\label{sec:liv-limits}

\begin{sidewaystable}[p]
  \centering
  \begin{tabular}{lccccccc}
    \hline\hline
    \multicolumn{1}{c}{Source}& Experiment & Limit on $\EQG^{(1)}$ &
    Limit on $\EQG^{(2)}$ & Distance & $\Delta t$ & $E_\mathrm{max}$ &
    Ref.\\
    \hline \\[-2.1ex]
    Crab   & VERITAS & $1.7 \cdot 10^{17} \U{GeV}$ & $7 \cdot 10^9 \U{GeV}$ &
      $2.2 \U{kpc}$ & $100 \U{\mu s}$ & $120 \U{GeV}$ &
      \cite{Zitzer:2013gka,Otte:2012tw}\\
    GRB090510 & Fermi/LAT & $9.1 \cdot 10^{19} \U{GeV}$ & 
      $1.3 \cdot 10^{11} \U{GeV}$ &
      $z=0.903$ & \multicolumn{2}{c}{combined methods} & 
      \cite{Vasileiou:2013vra}\\
    PKS 2155-304 & H.E.S.S. & $2.1 \cdot 10^{18} \U{GeV}$ &
      $6.4 \cdot 10^{10} \U{GeV}$ &
      $z = 0.116$ & \multicolumn{2}{c}{likelihood fit} &
      \cite{HESS:2011aa}\\
    PG 1553+113 & H.E.S.S., Fermi/LAT &
      $4.3 \cdot 10^{17} \U{GeV}$ & $2.1 \cdot 10^{10} \U{GeV}$ &
      $z = 0.49 \pm 0.04$ & \multicolumn{2}{c}{combined analysis} &
      \cite{Sanchez:2015yua}\\
    \hline \\[-2.1ex]
    HAWC Pulsar ref. & HAWC & $10^{17} \U{GeV}$ & 
      $9 \cdot 10^9 \U{GeV}$ & $2 \U{kpc}$ & $1 \U{ms}$ & 
      $500 \U{GeV}$ &\\
    HAWC GRB ref. & HAWC & $4.9 \cdot 10^{19}\U{GeV}$ &
      $1.1 \cdot 10^{11}\U{GeV}$ & $z = 1$ & $1 \U{s}$ &
      $100 \U{GeV}$ & \\
    \hline\hline
  \end{tabular}
  \caption{Compilation of the most stringent results on LIV published
    and the 
    potential of the HAWC observatory, based on the reference
    scenarios described in section~\protect\ref{sec:hawc-potential}.}
  \label{tab:limits}
\end{sidewaystable}

Gamma ray observatories have used observations of pulsars, GRBs, and
AGNs to set limits on the scale of QG. A compilation of results can be
found in \cite{Bolmont:2010np}, with additional results reported in
\cite{Otte:2012tw,Amelino-Camelia:2013naa}. The precise estimate of
the propagation delay is a non-trivial part of establishing limits on
LIV. A simplifying assumption one can make is that of simultaneous
emission of all photons. Doing this, one automatically over-estimates
the contribution of LIV to the spread of arrival times $\Delta t$,
since one neglects the contribution of astrophysical effects in the
source. Improvements in modeling the source and more sophisticated
analysis techniques can be used to obtain a better estimate of the
contribution of LIV and thereby set more stringent limits.

The current best limits on both the linear and quadratic term have
been set by Fermi/LAT's observation of GRB090510
\cite{Vasileiou:2013vra}. The limits set are $\EQG^{(1)} > 9.1 \cdot
10^{19}\U{GeV}$ and $\EQG^{(2)} > 1.3 \cdot 10^{11}\U{GeV}$.

The observation of the flare of PKS 2155-304 on MJD 53944 by H.E.S.S
provides the best limits on LIV derived from the observation of AGNs
\cite{HESS:2011aa}. The reported limit on the quadratic term of
$\EQG^{(2)} > 6.4 \cdot 
10^{10}\U{GeV}$ is within a factor of two of the curent best limit set
by Fermi/LAT. The joint analysis by H.E.S.S. and Fermi/LAT
of the flaring of PG 1553+113 in 
April 2012 \cite{Sanchez:2015yua} shows the potential of combining
results from multiple observatories, even though the limits are
slightly worse than those H.E.S.S obtained from PKS 2155-304.

The best limits derived from pulsars have been set by VERITAS'
observation of the pulsation of the Crab \cite{Aliu:2011zi}. The
limits reported are $\EQG^{(1)} > 1.7\cdot 10^{17}\U{GeV}$
\cite{Zitzer:2013gka} and $\EQG^{(2)} > 7 \cdot 10^9 \U{GeV}$
\cite{Otte:2012tw}.

\section{Potential of HAWC for setting limits on LIV}
\label{sec:hawc-potential}

We use the sources studied by other observatories in the analyses
quoted in section~\ref{sec:liv-limits} to construct reference
scenarios for the HAWC observatory 
and establish its potential to set limits on LIV.

The scenario for setting limits on LIV from GRBs is motivated by the
properties of the short burst GRB090510 \cite{2009Natur.462..331A}
and of the long burst GRB130427A \cite{Amelino-Camelia:2013naa}. Our
reference scenario is a short burst with $\Delta t = 1 \U{sec}$ at a
redshift of $z = 1$, with a maximum observed photon energy of
$E_\mathrm{max} = 100 \U{GeV}$, close to that of GRB130427A
\cite{Amelino-Camelia:2013naa}.  
Such a burst is detectable by HAWC if
it occurs in its field of view \cite{Abeysekara:2011yu}. The time
spread and redshift assumed in this scenario are clearly compatible
with observed GRBs used to set limits on LIV \cite{Abdo:2009zza,
  2009Natur.462..331A,Amelino-Camelia:2013naa}. 
In this scenario, it is possible for HAWC to set 
a limit of $4.9 \cdot 10^{19}\U{GeV}$ for
the linear term $\EQG^{(1)}$ and $1.1 \cdot 10^{11}\U{GeV}$ for
$\EQG^{(2)}$. Comparing these numbers with the limits reported in
section~\ref{sec:liv-limits} shows that HAWC can set competitive
limits even with very basic analysis techniques. Improvements to
modeling and analysis are expected to allow HAWC to set more stringent
limits. 

Our reference scenario for pulsars is motivated by the VERITAS
observation of the Crab pulsar at energies up to $120\U{GeV}$
\cite{Aliu:2011zi} 
and the limits on LIV derived from that observation
\cite{Zitzer:2013gka,Otte:2012tw}. We assume a 
time spread $\Delta t = 1\U{ms}$, a factor of 10 larger than used in
the VERITAS 
limit. This arbitrary degradation of the resolved time spread was
chosen since the potential of HAWC to observe pulsations, and
therefore details of the light curve, of the Crab Nebula has not yet
been demonstrated.  
The degenerated time resolution gets compensated by assuming an
increased maximum energy 
of $E_\mathrm{max} = 500\U{GeV}$ for a pulsar at a distance of $2 \U{kpc}$.
Assuming the observation of such a pulsar, HAWC could  establish limits
for the linear term $\EQG^{(1)}$ up to 
$10^{17}\U{GeV}$ and for the quadratic term $\EQG^{(2)}$ of up to 
$9 \cdot 10^9 \U{GeV}$. The assumption of an increase in
the observed $E_\mathrm{max}$ allows for an improved limit on the
quadratic term, despite the worsened limit on the time
difference. One has to assume that such a source, if it exists, will
also be observed by Imaging Air Cherenkov Telescopes (IACTs). 
It will remain to be seen if the HAWC's
ability to monitor a pulsar continuously will be able to compensate
for increased detail of observations by IACTs.

It is unlikely that HAWC will be able to improve on limits derived
from the observation of flaring AGNs by IACTs. The lower statistics of HAWC
results in less detail in light curves, which in turn results in a
less stringent estimate of $\Delta t$.

Of the sources considered here, one-shot transients like GRBs provide
the most favourable objects for setting stringent limits on
LIV\@. HAWC can reach 
higher in energy than satellites like Fermi and has a larger
probability to see a GRB in its field of view than an IACT. Combining
observations by HAWC with satellite based observations has the
potential to provide even better limits on LIV\@. Pulsars have the
advantage that they 
are long-lived and therefore reliably detectable, even though HAWC's
ability to observe pulsation still has to be established. The large
field of view and high duty cycle of HAWC results in a large
accumulated time on each source, which can make up for the detail in
the high statistics but time-limited observations by IACTs.

\section{Conclusions}
\label{sec:conclusions}

We developed scenarios establishing the potential of the HAWC
Observatory to set limits on LIV in models with modified speed of
propagation of photons. Using observations of GRBs, HAWC has the
potential to set competitive limits. So far, we consider only very simple
scenarios and analysis techniques. We expect to improve the potential
of HAWC in this area in the future.

\section*{Acknowledgments}
\footnotesize{
  We acknowledge the support from: the US National
  Science Foundation (NSF); the US Department of Energy Office of
  High-Energy Physics; the Laboratory Directed Research and
  Development (LDRD) program of Los Alamos National Laboratory;
  Consejo Nacional de Ciencia y Tecnolog\'{\i}a (CONACyT), Mexico
  (grants 260378, 55155, 105666, 122331, 132197, 167281, 167733); Red
  de F\'{\i}sica de Altas Energ\'{\i}as, Mexico; DGAPA-UNAM (grants
  IG100414-3, IN108713, IN121309, IN115409, IN111315); VIEP-BUAP
  (grant 161-EXC-2011); the University of Wisconsin Alumni Research
  Foundation; the Institute of Geophysics, Planetary Physics, and
  Signatures at Los Alamos National Laboratory; the Luc Binette
  Foundation UNAM Postdoctoral Fellowship program.
}

\bibliographystyle{JHEP}
\bibliography{icrc2015-1056}


\end{document}